\documentclass[aps,prd,twocolumn,showpacs,amsmath,amssymb]{revtex4-1}
\usepackage{amsmath}
\usepackage{graphicx}
\usepackage{subfigure}
\usepackage{epstopdf}
\usepackage{color}
\usepackage{multirow}
\usepackage{setspace}
\usepackage{overpic}
\usepackage{amssymb}
\usepackage[driverfallback=dvipdfm, bookmarksnumbered, pdfstartview=FitH,colorlinks,urlcolor=blue, citecolor=blue,linkcolor=blue] {hyperref}
\usepackage{lineno}
\usepackage{bm}
\usepackage{rotating}
\usepackage[utf8]{inputenc}
\usepackage{threeparttable}

\let\oldequation\equation
\let\oldendequation\endequation

\renewenvironment{equation}
  {\linenomathNonumbers\oldequation}
  {\oldendequation\endlinenomath}

\begin{document}

\title{\bf \boldmath
Measurements of the absolute branching fractions of hadronic $D$-meson decays involving kaons and pions
}

\author{
M.~Ablikim$^{1}$, M.~N.~Achasov$^{10,b}$, P.~Adlarson$^{69}$, S. ~Ahmed$^{15}$, M.~Albrecht$^{4}$, R.~Aliberti$^{29}$, A.~Amoroso$^{68A,68C}$, M.~R.~An$^{33}$, Q.~An$^{65,51}$, X.~H.~Bai$^{59}$
% ...
\\
\vspace{0.2cm}
(BESIII Collaboration)\\
\vspace{0.2cm} {\it
$^{1}$ Institute of High Energy Physics, Beijing 100049, People's Republic of China\\
% ...
$^{4}$ Bochum Ruhr-University, D-44780 Bochum, Germany\\
% ...
$^{10}$ G.I. Budker Institute of Nuclear Physics SB RAS (BINP), Novosibirsk 630090, Russia\\
% ...
$^{15}$ Helmholtz Institute Mainz, Staudinger Weg 18, D-55099 Mainz, Germany\\
% ...
$^{29}$ Johannes Gutenberg University of Mainz, Johann-Joachim-Becher-Weg 45, D-55099 Mainz, Germany\\
% ...
$^{33}$ Liaoning Normal University, Dalian 116029, People's Republic of China\\
% ...
$^{51}$ State Key Laboratory of Particle Detection and Electronics, Beijing 100049, Hefei 230026, People's Republic of China\\
% ...
$^{59}$ University of Jinan, Jinan 250022, People's Republic of China\\
% ...
$^{65}$ University of Science and Technology of China, Hefei 230026, People's Republic of China\\
% ...
$^{68}$ University of Turin and INFN, (A)University of Turin, I-10125, Turin, Italy; (B)University of Eastern Piedmont, I-15121, Alessandria, Italy; (C)INFN, I-10125, Turin, Italy\\
$^{69}$ Uppsala University, Box 516, SE-75120 Uppsala, Sweden\\
% ...
$^{b}$ Also at the Novosibirsk State University, Novosibirsk, 630090, Russia\\
% ...
}
}

\begin{abstract}
By analyzing an electron-positron collision data sample corresponding to an integrated luminosity of $2.93\,\rm fb^{-1}$ taken at the center-of-mass energy of 3.773 GeV with the BESIII detector, we obtain for the first time the absolute branching fractions for seven $D^0$ and $D^+$ hadronic decay modes and search for the hadronic decay $D^0\to K^0_S K^0_S\pi^0$ with much improved sensitivity. 
The results are
\begin{center}
${\mathcal B}(D^0\to K^0_S\pi^0\pi^0\pi^0     )=( 7.64\pm  0.30\pm  0.29)\times 10^{-3}$,\\
${\mathcal B}(D^0\to K^-\pi^+\pi^0\pi^0\pi^0  )=( 9.54\pm  0.30\pm  0.31)\times 10^{-3}$,\\
${\mathcal B}(D^0\to K^0_S\pi^+\pi^-\pi^0\pi^0)=(12.66\pm  0.45\pm  0.43)\times 10^{-3}$,\\
${\mathcal B}(D^+\to K^0_S\pi^+\pi^0\pi^0     )=(29.04\pm  0.62\pm  0.87)\times 10^{-3}$,\\
${\mathcal B}(D^+\to K^0_S\pi^+\pi^+\pi^-\pi^0)=(15.28\pm  0.57\pm  0.60)\times 10^{-3}$,\\
${\mathcal B}(D^+\to K^0_S\pi^+\pi^0\pi^0\pi^0)=( 5.54\pm  0.44\pm  0.32)\times 10^{-3}$,\\
${\mathcal B}(D^+\to K^-\pi^+\pi^+\pi^0\pi^0  )=( 4.95\pm  0.26\pm  0.19)\times 10^{-3}$,\\
${\mathcal B}({D^0\to K^0_S K^0_S\pi^0}) < 1.45 \times 10^{-4}$ at the 90\% confidence level.
\end{center}
Here the first uncertainties are statistical and the second ones systematic.
The newly studied decays greatly enrich the knowledge of the $D\to \bar K\pi\pi\pi$ and $D\to \bar K\pi\pi\pi\pi$ hadronic decays,
and open a bridge to access more two-body hadronic $D$ decays containing scalar, vector, axial and tensor mesons in the charm sector.
\end{abstract}


\maketitle

\oddsidemargin  -0.2cm
\evensidemargin -0.2cm

\section{Introduction}

Experimental investigations of hadronic $D$ decays can greatly aid our understanding of strong and weak interactions~\cite{bes3-white-paper,belle2-white-paper,lhcb-white-paper}.
For example, studies of hadronic $D$ decays provide a way to explore the effects of $D^0-\bar D^0$ mixing and charge-parity ($CP$) violation, which are key to understanding the asymmetry between matter and anti-matter in the universe~\cite{asymmetry-adams}.
In addition, the improved knowledge of the strong phase difference in various hadronic decays of neutral $D$ mesons provides key information needed to extract the Cabibbo-Kobayashi-Maskawa (CKM) triangle angle of $\gamma$ in $B$ physics, which is crucial to test CKM matrix unitarity.
In addition, amplitude analyses of multi-body hadronic $D$ decays help to access quasi-two-body hadronic $D$ decays and to extract chiral structures of weak interaction in analogy to the multi-body decay processes explored in~\cite{Altmannshofer2011,Beneke2014}.
Moreover, combining amplitude analysis results with the precisely measured branching fractions of these same hadronic $D$ decays yields the branching fractions of two-body hadronic $D$ decays, which are important to explore the phenomenon of quark SU(3)-flavor symmetry breaking~\cite{ref5,theory_1,theory_2,chenghy1,yufs}.

Since the discovery of $D$ mesons in the 1970s, hadronic $D$ decays have been investigated extensively and precisely~\cite{pdg2020}. However, some multi-body Cabibbo-favored decays, e.g.,
$D^0\to K^0_S\pi^0\pi^0\pi^0$,
$K^-\pi^+\pi^0\pi^0\pi^0$,
$K^0_S\pi^+\pi^-\pi^0\pi^0$,
$D^+\to K^0_S\pi^+\pi^0\pi^0$,
$K^0_S\pi^+\pi^0\pi^0\pi^0$,
$K^-\pi^+\pi^+\pi^0\pi^0$, and
$K^0_S\pi^+\pi^+\pi^-\pi^0$,
remain unmeasured. Experimental studies of these decays are challenging mainly due to high background, low efficiency, and poor resolution.
In this paper, we report on the measurements of the absolute branching fractions for these multi-body decays by analyzing the $e^+e^-$ collision data sample corresponding to an integrated luminosity of 2.93~fb$^{-1}$~\cite{lum_bes31,lum_bes32} collected at the
center-of-mass energy of $\sqrt s=$ 3.773~GeV with the BESIII detector.
Throughout this paper, charge conjugate processes are always implied.

\section{BESIII detector and Monte Carlo simulation}
The BESIII detector is a magnetic spectrometer~\cite{BESIII} located at the Beijing Electron
Positron Collider (BEPCII)~\cite{Yu:IPAC2016-TUYA01}. The
cylindrical core of the BESIII detector consists of a helium-based
 multilayer drift chamber (MDC), a plastic scintillator time-of-flight
system (TOF), and a CsI (Tl) (TI doped CsI crystal) electromagnetic calorimeter (EMC),
which are all enclosed in a superconducting solenoidal magnet
providing a 1.0~T magnetic field. The solenoid is supported by an
octagonal flux-return yoke with resistive-plate counter muon-identifier modules interleaved with steel. The acceptance of
charged particles and photons is 93\% over $4\pi$ solid angle. The
charged-particle momentum resolution at $1~{\rm GeV}/c$ is
$0.5\%$, and the resolution of the specific ionization energy loss ($dE/dx$) is $6\%$ for the electrons
from Bhabha scattering. The EMC measures photon energies with a
resolution of $2.5\%$ ($5\%$) at $1$~GeV in the barrel (end cap)
region. The time resolution of the TOF barrel part is 68~ps, while
that of the end cap part is 110~ps.

Simulated samples, produced with the {\sc geant4}-based~\cite{geant4} Monte Carlo (MC) package including the geometric description of the BESIII detector and the
detector response, are used to determine the detection efficiency
and to estimate the backgrounds. The simulation includes the beam-energy spread and initial-state radiation in the $e^+e^-$
annihilations modeled with the generator {\sc kkmc}~\cite{kkmc1}.
The inclusive MC samples consist of the production of $D\bar{D}$
pairs, the non-$D\bar{D}$ decays of the $\psi(3770)$, the initial-state radiation
production of the $J/\psi$ and $\psi(3686)$ states, and the
continuum processes.
The known decay modes are modeled with {\sc
evtgen}~\cite{evtgen1} using the branching fractions taken from the
Particle Data Group (PDG)~\cite{pdg2020}, and the remaining unknown decays
from the charmonium states are modeled with {\sc
lundcharm}~\cite{lundcharm1,lundcharm2}. The final-state radiation
from charged final-state particles is incorporated with the {\sc
photos} package~\cite{photos}.

\section{Measurement Method}
The $D^0\bar D^0$ or $D^+D^-$ pairs are produced without any additional hadron in $e^+e^-$ annihilations at $\sqrt s=3.773$ GeV. This process offers a clean environment to measure the branching fractions of hadronic $D$ decays with the double-tag method~\cite{double-tag}.
The single-tag candidate events are selected by reconstructing a $\bar D^0$ or $D^-$ in the following hadronic final states:
$\bar D^0 \to K^+\pi^-$, $K^+\pi^-\pi^0$, $K^+\pi^-\pi^-\pi^+$, and
$D^- \to K^{+}\pi^{-}\pi^{-}$,
$K^0_{S}\pi^{-}$, $K^{+}\pi^{-}\pi^{-}\pi^{0}$, $K^0_{S}\pi^{-}\pi^{0}$, $K^0_{S}\pi^{+}\pi^{-}\pi^{-}$, $K^{+}K^{-}\pi^{-}$.
The event in which a signal candidate is selected in the presence of a single-tag (ST) $\bar D$ meson
is called a double-tag (DT) event.
The branching fraction for the signal decay is determined by
\begin{equation}
\label{eq:br}
{\mathcal B}_{{\rm sig}} = N^{\rm net}_{\rm DT}/(N^{\rm tot}_{\rm ST}\cdot\epsilon_{{\rm sig}}),
\end{equation}
where $N^{\rm tot}_{\rm ST}=\sum_i N_{{\rm ST}}^i$ and $N^{\rm net}_{\rm DT}$ are the total single-tag yield and signal yield in data, respectively, in which $N_{{\rm ST}}^i$ is the single-tag yield for the tag mode $i$. For the signal decays involving $K^0_S$ meson(s) in the final states, $N^{\rm net}_{\rm DT}$ is
the net signal yield after removing the peaking background, which is dominated by the corresponding non-$K^0_S$ decays. The yield of the peaking background is also obtained from the fit with the double-tag method.
For the other signal decays, the variable only corresponds to
the fitted double-tag yields. Further details are described in Sec. 6.
Here, $\epsilon_{{\rm sig}}$ is the efficiency of detecting the signal $D$ decay, averaged over all tag modes $i$, which is given by
\begin{equation}
\label{eq:eff}
\epsilon_{{\rm sig}} = \sum_i (N^i_{{\rm ST}}\cdot\epsilon^i_{{\rm DT}}/\epsilon^i_{{\rm ST}})/N^{\rm tot}_{\rm ST},
\end{equation}
where $\epsilon^i_{{\rm ST}}$ and $\epsilon^i_{{\rm DT}}$ are the efficiencies of detecting single-tag and double-tag candidates in the tag mode $i$, respectively.

The measurements of the branching fractions of neutral, self-conjugate $D$ decays have to be corrected for the effects of {quantum correlation (QC) existing in the data but not being implemented in the MC simulation.} For each neutral $D$ decay, the $CP$-even component is estimated by the $CP$-even tag $D^0\to K^+K^-$
and the $CP$-odd tag $D^0\to K^0_S\pi^0$.
Using the same method as described in Ref.~\cite{QC-factor}
and the parameters quoted from Refs.~\cite{R-ref1,R-ref2,R-ref3},
we find the correction factors ($f_{\rm QC}$) that account for the QC effect on the measured branching fractions to be {$1.081\pm{0.007_{\rm stat}}$ and $0.956\pm{0.006_{\rm stat}}$ for
$D^0\to K^0_S\pi^0\pi^0\pi^0$ and $D^0\to K^0_S\pi^+\pi^-\pi^0\pi^0$}, respectively.  Here, $f_{\rm QC}$ multiplies the naively extracted branching fractions.

\section{Event selection}

The selection criteria of $K^\pm$, $\pi^\pm$, $K^0_S$, and $\pi^0$ are the same as those used in the analyses presented in
Refs.~\cite{epjc76,cpc40,bes3-pimuv}.
All charged tracks, except those originating from $K^0_{S}$ decays, are required to have a polar angle $\theta$ with respect to the beam direction
within the MDC acceptance $|\rm{\cos\theta}|<0.93$,  and a distance of closest approach to the interaction point within 10~cm along the beam direction
and within 1~cm in the plane perpendicular to the beam direction. Particle identification (PID) for charged pions and kaons is performed by
exploiting TOF information and the $dE/dx$ measured by the MDC.
The confidence levels for pion and kaon hypotheses ($CL_{\pi}$ and $CL_{K}$) are calculated. Kaon and pion candidates are required to
satisfy $CL_{K}>CL_{\pi}$ and $CL_{\pi}>CL_{K}$, respectively.

The $K^0_S$ candidates are reconstructed from two oppositely charged tracks which are assigned as pions with no PID criteria applied.
These charged tracks must satisfy $|\rm{\cos\theta}|<0.93$. In addition, due to the long lifetime of the $K^0_S$  meson,
there is a less stringent criterion on the distance of closest approach to the interaction point in the beam direction of less than 20~cm and no requirement on the
distance of closest approach in the plane transverse to the beam direction. Furthermore, the $\pi^+\pi^-$ pairs are constrained to originate from a common vertex and their invariant mass is required to be within $(0.486,0.510)~{\rm GeV}/c^2$,
which corresponds to about three times the fitted resolution around the $K^0_S$ nominal mass. The decay length of the $K^0_S$ candidate is required to be
greater than two standard deviations of the vertex resolution away from the interaction point.

The $\pi^0$ candidate is reconstructed via its $\gamma\gamma$ decay. The photon candidates are selected using the information from the EMC showers.
It is required that each EMC shower starts within 700~ns of the event start time and its energy is greater than 25 (50)~MeV in the barrel (end cap)
region of the EMC~\cite{BESIII}. The energy deposited in the neighboring TOF counters is included to improve the reconstruction efficiency and energy resolution as the correctness of the shower energy. The opening angle between the candidate shower and the nearest charged track must be greater than $10^{\circ}$. The $\gamma\gamma$ pair is taken as a $\pi^0$ candidate
if its invariant mass is within $(0.115,\,0.150)$\,GeV$/c^{2}$. To improve the resolution, a kinematic fit constraining the $\gamma\gamma$
invariant mass to the $\pi^{0}$ nominal mass~\cite{pdg2020} is imposed on the selected photon pair.

\section{Yields of single-tag $\bar D$ mesons}

To select $\bar D^0\to K^+\pi^-$ candidates, the backgrounds from cosmic rays and Bhabha events are rejected by using the same requirements described in Ref.~\cite{deltakpi}.
In the selection of $\bar D^0\to K^+\pi^-\pi^-\pi^+$ candidates, the $\bar D^0\to K^0_SK^\pm\pi^\mp$ decays are suppressed by requiring the mass of all $\pi^+\pi^-$ pairs to
be outside $(0.483,0.513)$~GeV/$c^2$.

The tagged $\bar D$ mesons are identified using two variables, namely the energy difference
\begin{equation}
\Delta E_{\rm tag} \equiv E_{\rm tag} - E_{\rm b},
\label{eq:deltaE}
\end{equation}
and the beam-constrained mass
\begin{equation}
M_{\rm BC}^{\rm tag} \equiv \sqrt{E^{2}_{\rm b}-|\vec{p}_{\rm tag}|^{2}}.
\label{eq:mBC}
\end{equation}
Here, $E_{\rm b}$ is the beam energy,
$\vec{p}_{\rm tag}$ and $E_{\rm tag}$ are the momentum and energy of
the $\bar D$ candidate in the rest frame of $e^+e^-$ system, respectively.
For each tag mode, if there are multiple candidates (about 10\% of the selected events) in an event,
only the one with the least $|\Delta E_{\rm tag}|$ is kept.
The tagged $\bar D$ candidates are required to satisfy
$\Delta E_{\rm tag}\in(-55,40)$\,MeV for the tag modes containing $\pi^0$ in the final states
and $\Delta E_{\rm tag}\in(-25,25)$\,MeV for the other tag modes, due to differing resolutions.

\begin{figure}[htp]
  \centering
\includegraphics[width=1.0\linewidth]{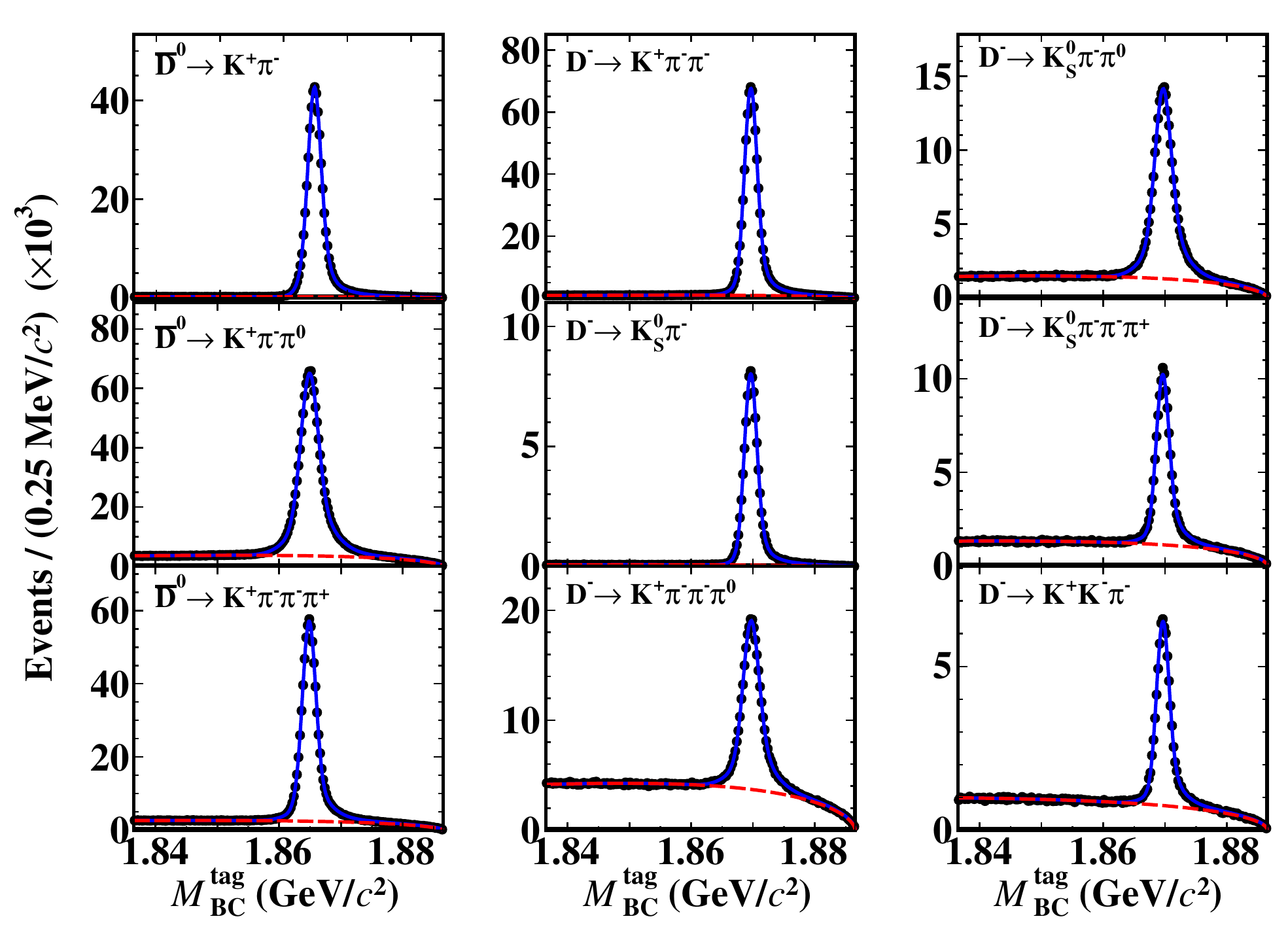}
  \caption{\small
Fits to the $M_{\rm BC}$ distributions of
the single-tag $\bar D^0$ (left column) and $D^-$ (middle and right columns) candidates,
where the points with error bars are data,
the blue solid and red dashed curves are the fit results
and the fitted backgrounds, respectively.}
\label{fig:datafit_MassBC}
\end{figure}

To extract the yields of single-tag $\bar D$ mesons for individual tag modes, binned maximum-likelihood fits are performed on the $M_{\rm BC}^{\rm tag}$
distributions of the single-tag candidates, following Refs.~\cite{epjc76,cpc40,bes3-pimuv}.
In the fits, the $\bar D$ signal is modeled by an MC-simulated shape convolved with
a double-Gaussian function describing the resolution difference between data and MC simulation.
The combinatorial background shape is described by an ARGUS function~\cite{ARGUS}
defined as $c_{M^{\rm tag}_{\rm BC}}(M^{\rm tag}_{\rm BC};E_{\rm end},\xi_{M^{\rm tag}_{\rm BC}})=A_{M^{\rm tag}_{\rm BC}}\cdot {M^{\rm tag}_{\rm BC}}\cdot \sqrt{1 - \frac {{M^{\rm tag}_{\rm BC}}^2}{E^2_{\rm end}/c^4}} \cdot \exp\left[\xi_{M^{\rm tag}_{\rm BC}} \left(1-\frac {{M^{\rm tag}_{\rm BC}}^2}{E^2_{\rm end}/c^4}\right)\right]$,
where
$E_{\rm end}$ is an endpoint fixed at 1.8865 GeV (corresponding to the beam energy), $A_{M^{\rm tag}_{\rm BC}}$ is a normalization factor, and $\xi_{M^{\rm tag}_{\rm BC}}$ is a free parameter.
The resulting fits to the $M_{\rm BC}$
distributions for various tag modes are shown in
Fig.~\ref{fig:datafit_MassBC}.
The total yields of the single-tag $\bar D^0$  and $D^-$ mesons in data are $(232.8\pm0.2)\times 10^4$ and
$(155.8\pm0.2)\times 10^4$, respectively, where the uncertainties are statistical only.

\section{Yields of double-tag events}

The signal $D$ decays are selected
by using the remaining tracks and showers that have not been used to reconstruct the single-tag $\bar D$ candidates.  Charged $D$ signal candidates
must have charge opposite to the tag; for neutral $D$ signal
candidates must have oppositely-charged kaons in the cases where both
kaons are charged.
For $D^0\to K^0_S\pi^+\pi^-\pi^0\pi^0$ and $D^+\to K^0_S\pi^+\pi^+\pi^-\pi^0$,
to suppress backgrounds from $D^0\to K^0_SK^0_S(\to \pi^+\pi^-)\pi^0\pi^0$ and $D^+\to K^0_SK^0_S(\to \pi^+\pi^-)\pi^+\pi^0$ decays, the
$\pi^+\pi^-$ invariant masses are required to be outside $(0.468,0.528)$~GeV/$c^2$. The signal decays that have two or more $\pi^0$s can have background contamination from a corresponding decay where the $\pi^0\pi^0$ pair originates from a $K^0_S$ meson. The rate of these backgrounds is typically small and are estimated with the known branching fractions taken from the PDG~\cite{pdg2020} except for the background of $D^0\to K^0_SK^0_S(\to \pi^0\pi^0)\pi^0$ for $D^0\to K^0_S\pi^0\pi^0\pi^0$. The upper limit on the branching fraction of $D^0\to K^0_SK^0_S\pi^0$ is re-estimated via $K^0_S (\to \pi^+\pi^-) K^0_S (\to \pi^+\pi^-)\pi^0$.
The data obtained at BESIII allow for a more accurate estimate than the current PDG limit~\cite{pdg2020}.
Since these peaking backgrounds are small, a veto is not applied on $M_{\pi^0\pi^0}$ due to a significant reduction of the signal efficiency.
The expected yields of these peaking backgrounds are fixed in our fit to $M^{\rm tag}_{\rm BC}$ vs.~$M^{\rm sig}_{\rm BC}$.

The signal $D$ mesons are identified using the energy difference $\Delta E_{\rm sig}$
and the beam-constrained mass $M_{\rm BC}^{\rm sig}$, which are calculated with the ``sig'' analogues of the ``tag'' Eqs.~(\ref{eq:deltaE}) and (\ref{eq:mBC}).
For each signal mode, if there are multiple candidates in an event, only the one with the smallest $|\Delta E_{\rm sig}|$ is kept.
The signal decays are required to satisfy the mode-dependent $\Delta E_{\rm sig}$ requirements,
as shown in the second column of Table~\ref{tab:DT}.

\begin{figure}[htp]
  \centering
  \includegraphics[width=1.0\linewidth]{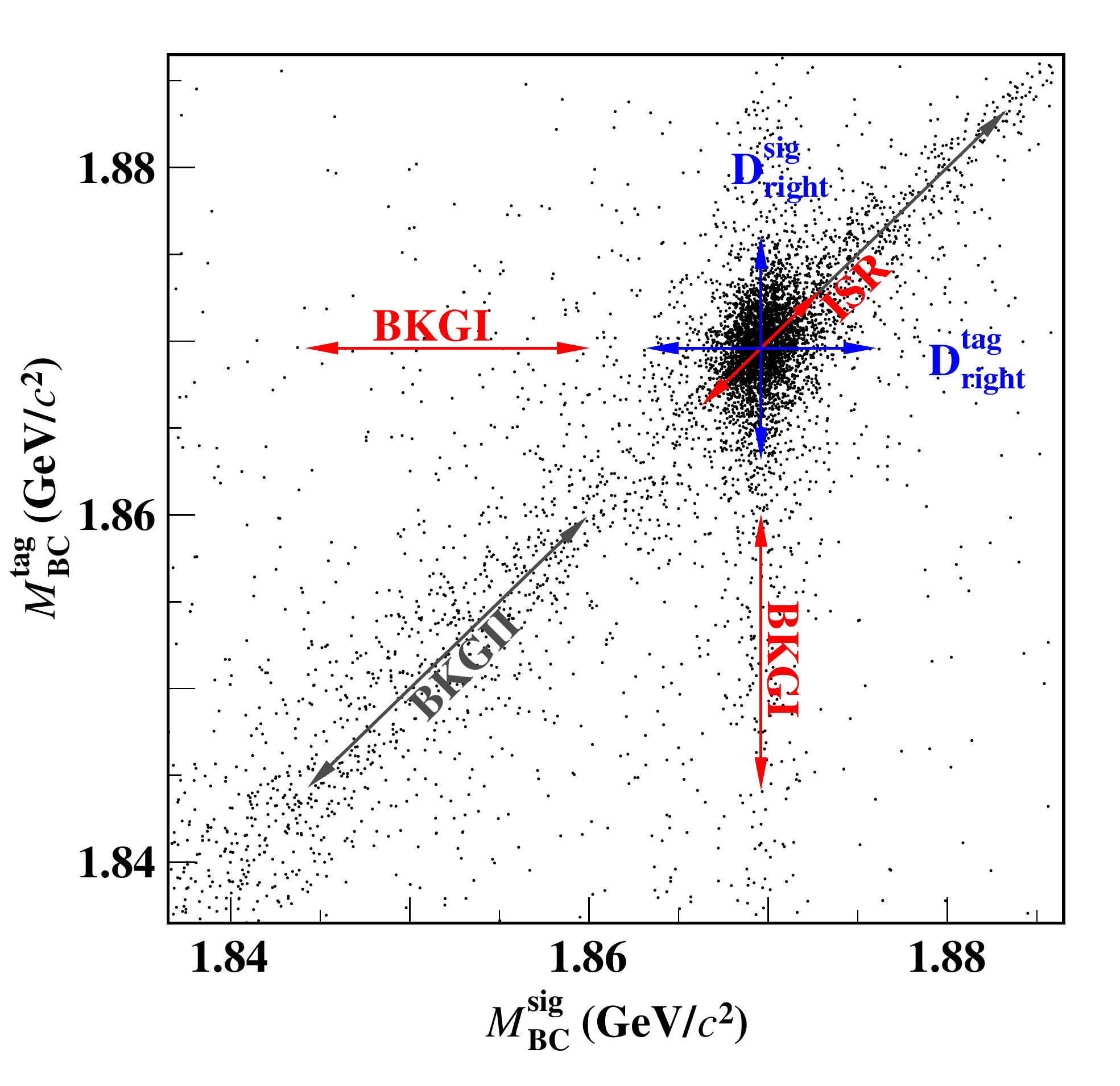}
  \caption{
    Distribution of $M_{\rm BC}^{\rm tag}$
    versus $M_{\rm BC}^{\rm sig}$ of the accepted candidates for $D^+\to K^0_S\pi^+\pi^0\pi^0$ versus all $D^-$ tag modes in data.
    Here, ISR denotes the signal spreading along the diagonal direction. The $D^{\rm sig}_{\rm right}$ and $D^{\rm tag}_{\rm right}$ denotes the signal spreading around $M_{\rm BC}^{\rm sig} = M_{D}$ and $M_{\rm BC}^{\rm tag} = M_{D}$.
}
\label{fig:mBC2D}
\end{figure}

Figure~\ref{fig:mBC2D} shows the $M_{\rm BC}^{\rm tag}$
versus $M_{\rm BC}^{\rm sig}$ distribution of the accepted double-tag candidates in data.
The signal events concentrate around $M_{\rm BC}^{\rm tag} = M_{\rm BC}^{\rm sig} = M_{D}$,
where $M_{D}$ is the $D$ nominal mass~\cite{pdg2020}.
The events with correctly reconstructed $D$ ($\bar D$) and incorrectly
reconstructed $\bar D$ ($D$), defined as BKGI, are spread along the lines around
$M_{\rm BC}^{\rm tag} = M_{D}$ or $M_{\rm BC}^{\rm sig} = M_{D}$.
The events smeared along the diagonal, defined as BKGII,
are mainly from the $e^+e^- \to q\bar q$ processes and incorrectly reconstructed $D\bar{D}$.
The events with uncorrelated and incorrectly reconstructed $D$ and $\bar D$, defined as BKGIII,
disperse in the whole allowed kinematic region.

For each signal $D$ decay mode, the yield of double-tag events ($N^{\rm fit}_{\rm DT}$) is obtained from a two-dimensional (2D) binned maximum-likelihood
fit~\cite{cleo-2Dfit} on the $M_{\rm BC}^{\rm tag}$ versus $M_{\rm BC}^{\rm sig}$ distribution of the accepted candidates. In the fit, the probability
density functions (PDFs) of signal, BKGI, BKGII, and BKGIII are constructed as
\begin{itemize}
\item
signal: $a(x,y)$,
\item
BKGI: $b(x)\cdot c_y(y;E_{\rm b},\xi_{y}) + b(y)\cdot c_x(x;E_{\rm b},\xi_{x})$,
\item
BKGII: $c_z(z;\sqrt{2}E_{\rm b},\xi_{z}) \cdot g(k)$, and
\item
BKGIII: $c_x(x;E_{\rm b},\xi_{x}) \cdot c_y(y;E_{\rm b},\xi_{y})$,
\end{itemize}
respectively.
Here, $x=M_{\rm BC}^{\rm sig}$, $y=M_{\rm BC}^{\rm tag}$, $z=(x+y)/\sqrt{2}$, and $k=(x-y)/\sqrt{2}$.
The PDFs of signal $a(x,y)$, $b(x)$, and $b(y)$ are described by the corresponding MC-simulated shapes and $c_f(f;E_{\rm end},\xi_f)$ is an ARGUS function~\cite{ARGUS} defined above,
where $f$ denotes $x$, $y$, or $z$; $E_{\rm b}$ is fixed at 1.8865 GeV. The signal shape $a(x, y)$ is also convolved with a 2D Gaussian function. The PDF $g(k)$ is a Gaussian function with mean of zero and standard deviation parametrized by $\sigma_k=\sigma_0 \cdot(\sqrt{2}E_{\rm b}/c^2-k)^p$,
where $\sigma_0$ and $p$ are fit parameters.

The signal decays with one $K^0_S (\to \pi^+\pi^-)$ have background contamination from corresponding decays which have combinatorial $\pi^+\pi^-$ pairs that satisfy the $K^0_S$ selection criteria. They form peaking backgrounds around $M_D$ in the $M_{\rm BC}^{\rm sig}$ distributions. This kind of peaking background is estimated by selecting events in the one-dimensional (1D) $K^0_S$ sideband region of
$(0.454,0.478)\cup(0.518,0.542)~{\rm GeV}/c^2$.

\begin{figure}[htp]
  \centering
  \includegraphics[width=1.0\linewidth]{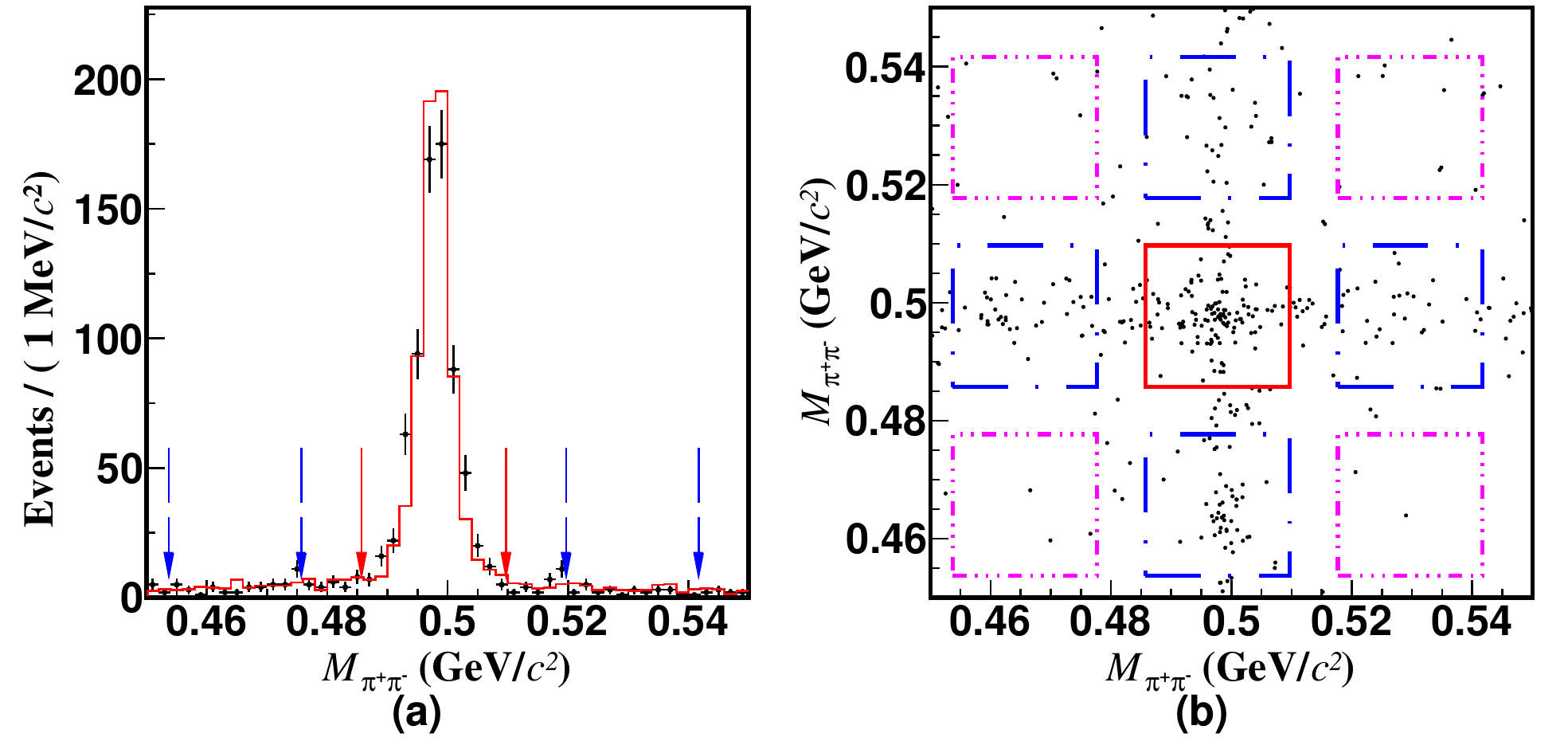}
\caption{\small
(a)~$\pi^+\pi^-$ invariant-mass distributions of the $D^0\to K^0_S\pi^0\pi^0\pi^0$ candidate events
of data (points with error bars) and inclusive MC sample (histogram).
Pairs of the red solid~(blue dashed) arrows denote the $K^0_S$ signal~(sideband) regions.
(b)~Distribution of $M_{\pi^+\pi^-(1)}$ versus $M_{\pi^+\pi^-(2)}$ for the $D^0\to K^0_SK^0_S\pi^0$ candidate events in data.
The red solid box denotes the 2D signal region.
Blue dashed~(pink dot-dashed) boxes indicate the 2D sideband 1~(2) regions.
}\label{fig:mks}
\end{figure}

Since there are two $K^0_S$ mesons in $D^0\to K^0_SK^0_S\pi^0$, 2D signal and sideband regions are used.
The 2D $K^0_S$ signal region is defined as the square region with both $\pi^+\pi^-$ combinations lying in the $K^0_S$ signal regions.
{The 2D $K^0_S$ sideband 1 regions are defined as the square regions with one $\pi^+\pi^-$ combination located in the 1D $K^0_S$ sideband regions and the other in the 1D $K^0_S$ signal region.
The sideband 2 regions are defined as the square regions with both $\pi^+\pi^-$ combinations located in the 1D $K^0_S$ sideband regions.}
Figure~\ref{fig:mks} shows 1D and 2D $\pi^+\pi^-$ invariant-mass distributions as well as the $K^0_S$ signal and sideband regions.

For the signal decays involving $K^0_S$ meson(s) in the final states,
the net yields of DT events are calculated by subtracting the sideband contribution from the DT fitted yield by
\begin{equation}
\label{eq:1}
N^{\rm net}_{\rm DT}
 = N^{\rm fit}_{\rm DT} + \sum^N_{i} \left [\left (-\frac{1}{2} \right )^{i} N^{\rm fit}_{{\rm sid}i} \right ],
\end{equation}
where $N^{\rm fit}_{\rm DT}$ and $N^{\rm fit}_{{\rm sid}i}$
are the fitted $D$ yields in the 1D or 2D signal region and sideband $i$ region, respectively, where $i$ runs from 1. This relation has been verified by a large MC sample. Here, $N=1$ for the decays with one $K^0_S$ meson, while $N=2$ for the decays with two $K^0_S$ mesons.
The combinatorial $\pi^+\pi^-$ backgrounds are assumed to be uniformly distributed and double-counting is avoided by subtracting sideband 2 yields from sideband 1
yields appropriately. For the other signal decays, the net yields of double-tag events are $N^{\rm fit}_{\rm DT}$.

\begin{figure}[htb]
  \centering
\includegraphics[width=1.0\linewidth]{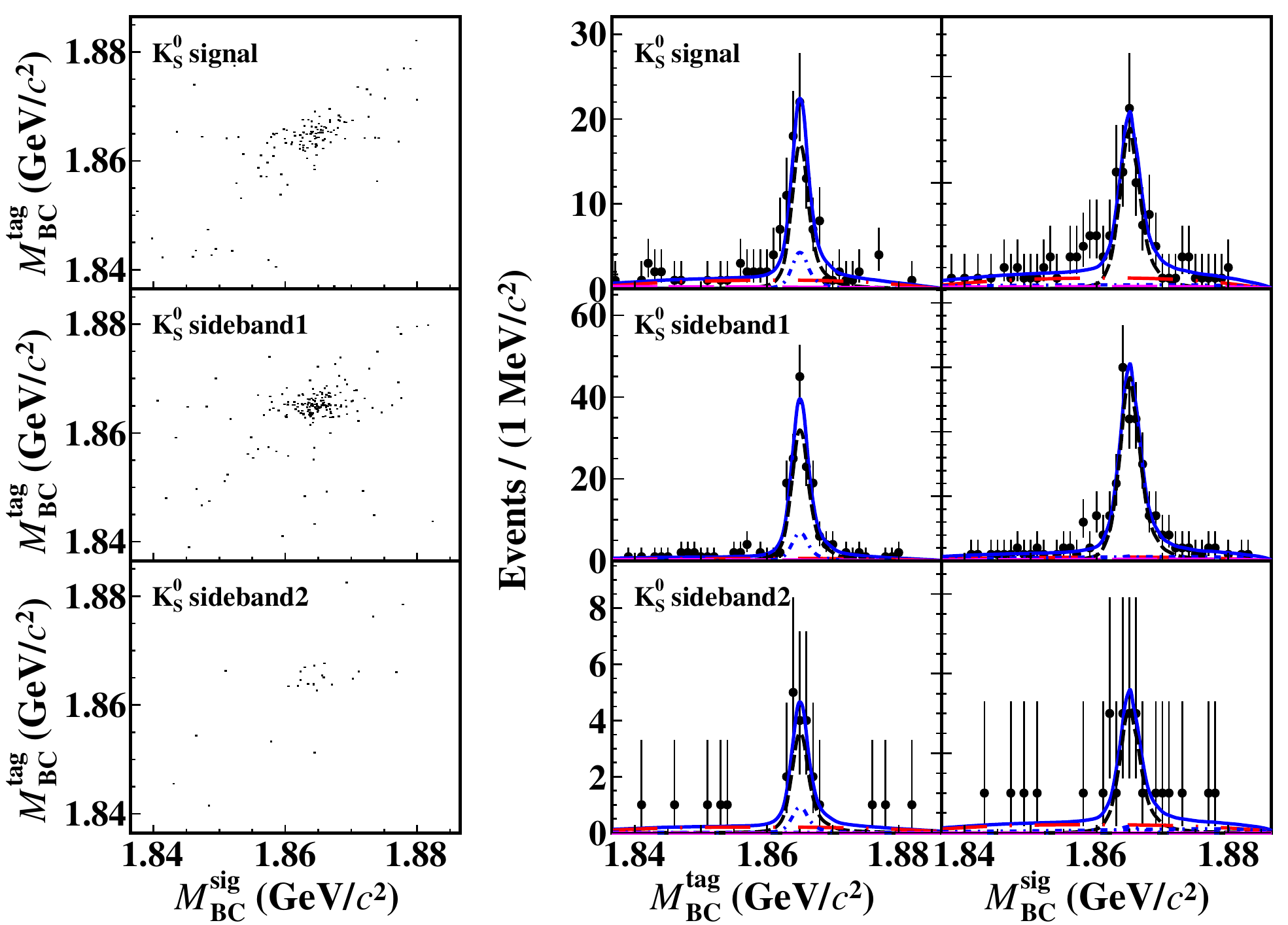}
  \caption{Scatter plots (left) and projections of $M_{\rm BC}^{\rm tag}$ and $M_{\rm BC}^{\rm sig}$ (right) for the 2D fits
on the candidate events for $D^0\to K^0_SK^0_S\pi^0$.
In the projections, the dots with error bars are data, and the blue solid curves are the total fit results.
The black dotted curves are the fitted signal, the blue dot-dashed curves are the BKGI, the red dot-long-dashed curves are the BKGII and the pink long-dashed curves are the BKGIII.}
\label{2Dfit_77}
\end{figure}

To obtain a more reliable peaking background yield from $D^0\to K^0_S(\to \pi^+\pi^-)K^0_S(\to \pi^0\pi^0)\pi^0$  in the study of
$D^0\to K^0_S \pi^0\pi^0\pi^0$, we have re-estimated the branching fraction of $D^0\to K^0_S K^0_S \pi^0$ via $K^0_S(\to \pi^+\pi^-) K^0_S(\to \pi^+\pi^-)\pi^0$.
Simultaneous two-dimensional maximum-likelihood fits are performed on the candidates for $D^0\to K^0_S(\to \pi^+\pi^-) K^0_S(\to \pi^+\pi^-)\pi^0$
in the 2D $K^0_S$ signal, sideband 1 and sideband 2 regions, as shown in Fig.~\ref{2Dfit_77}.

\begin{figure}[htbp]
  \centering
\includegraphics[width=1.0\linewidth]{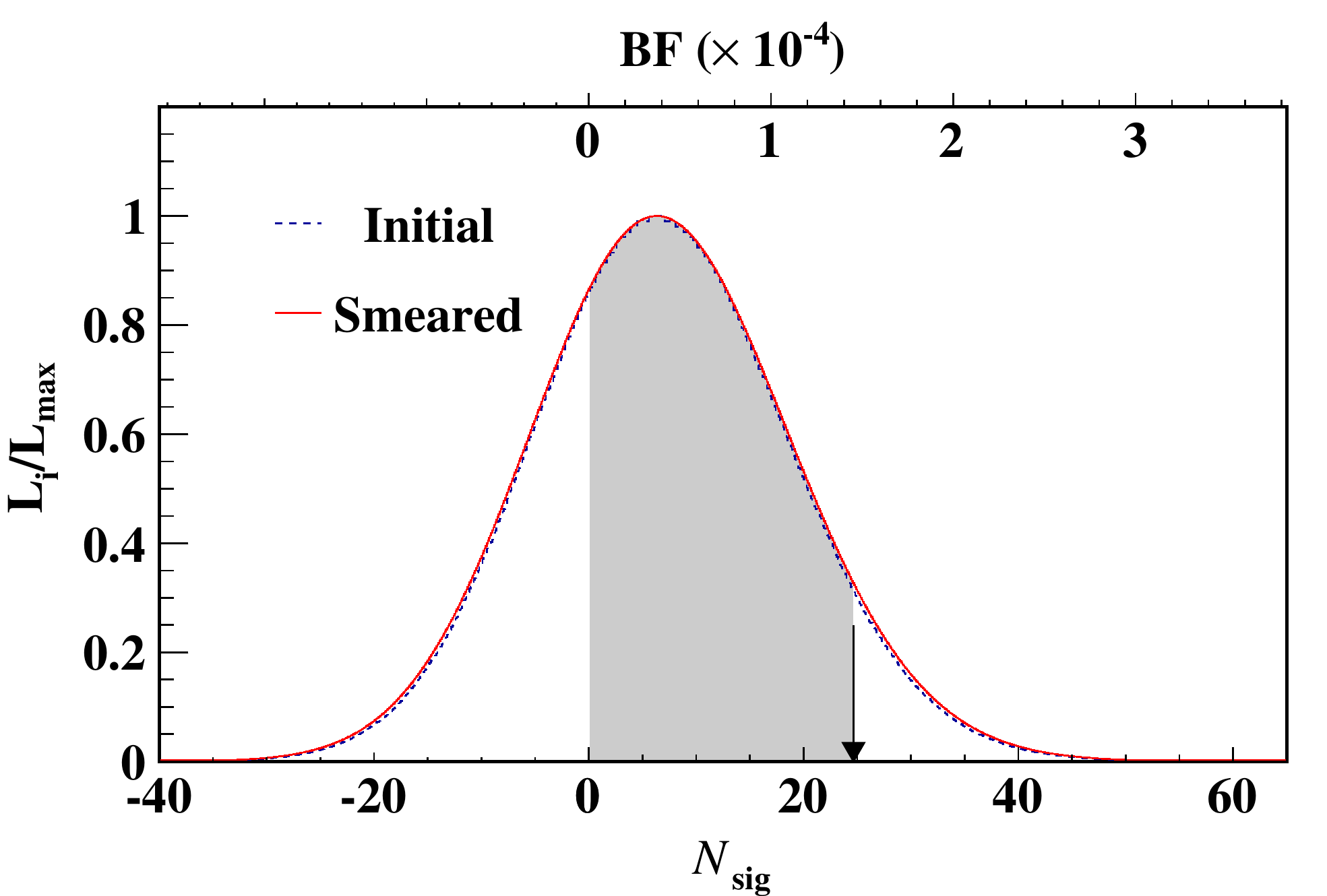}
  \caption{
Distribution of likelihood versus the assumed signal yield or branching fraction for $D^0\to K^0_S K^0_S\pi^0$. Here,  $\rm L_{max}$ denotes the maximum likelihood obtained from the fit. The results obtained with and without incorporating the systematic uncertainties are shown as the red solid and black dashed curves, respectively.
The black arrow shows the result corresponding to the 90$\%$ confidence level.}
\label{scan}
\end{figure}

\begin{figure}[htbp]
  \centering
\includegraphics[width=1.0\linewidth]{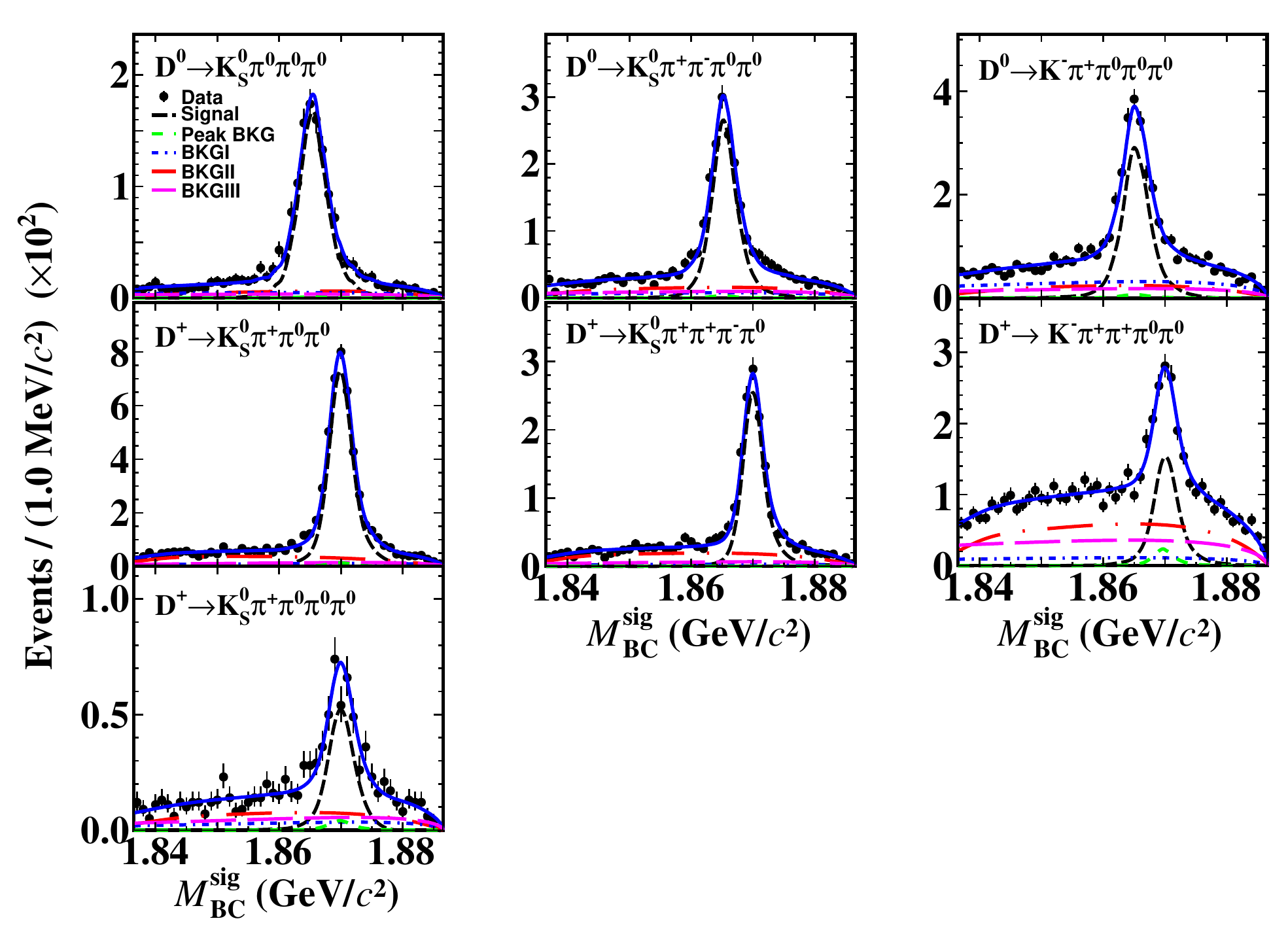}
\includegraphics[width=1.0\linewidth]{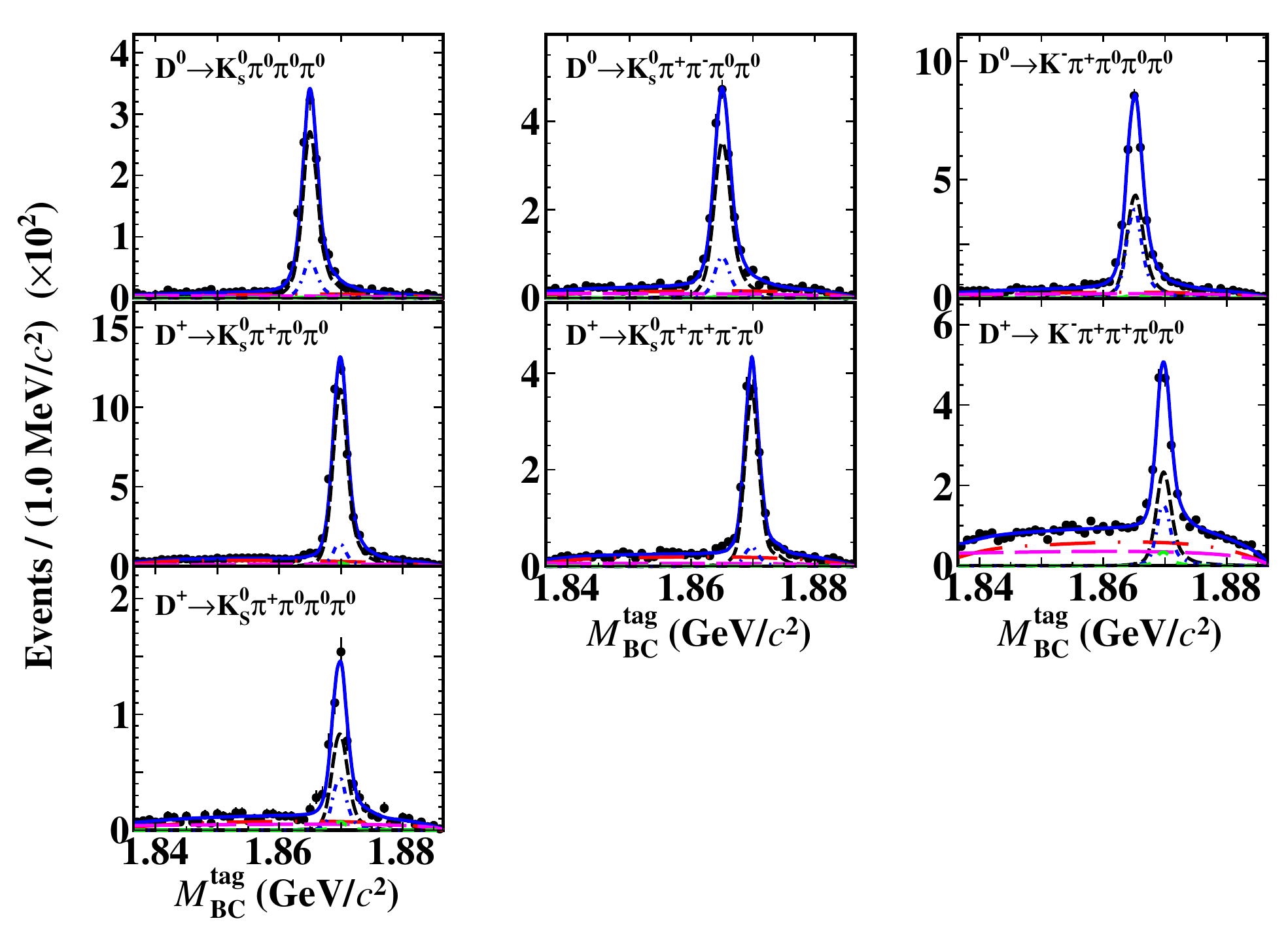}
  \caption{\small
Projections of $M^{\rm tag}_{\rm BC}$ and
$M^{\rm sig}_{\rm BC}$ distributions for the 2D fits to the double-tag candidate events with all $\bar D^0$ or $D^-$ tags.
Data are shown as points with error bars.
Blue solid, black dotted, blue dot-dashed, red dot-long-dashed, pink long-dashed and green dashed curves denote the overall fit results,
signal, BKGI, BKGII, BKGIII and peaking background components (see text), respectively.
}
\label{fig:2Dfit}
\end{figure}
\begin{figure}[htb]
  \centering
\includegraphics[width=1.0\linewidth]{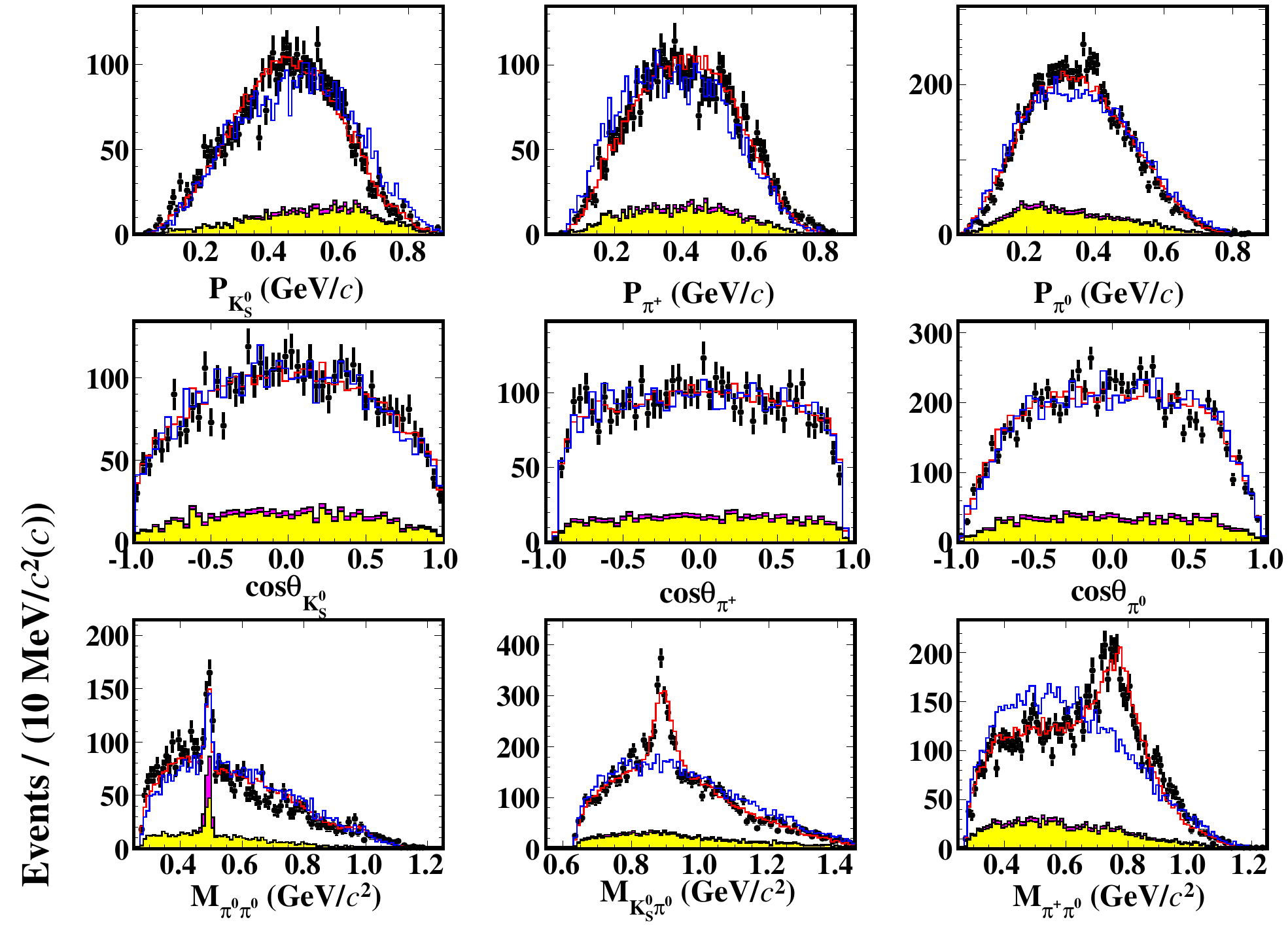}
\includegraphics[width=1.0\linewidth]{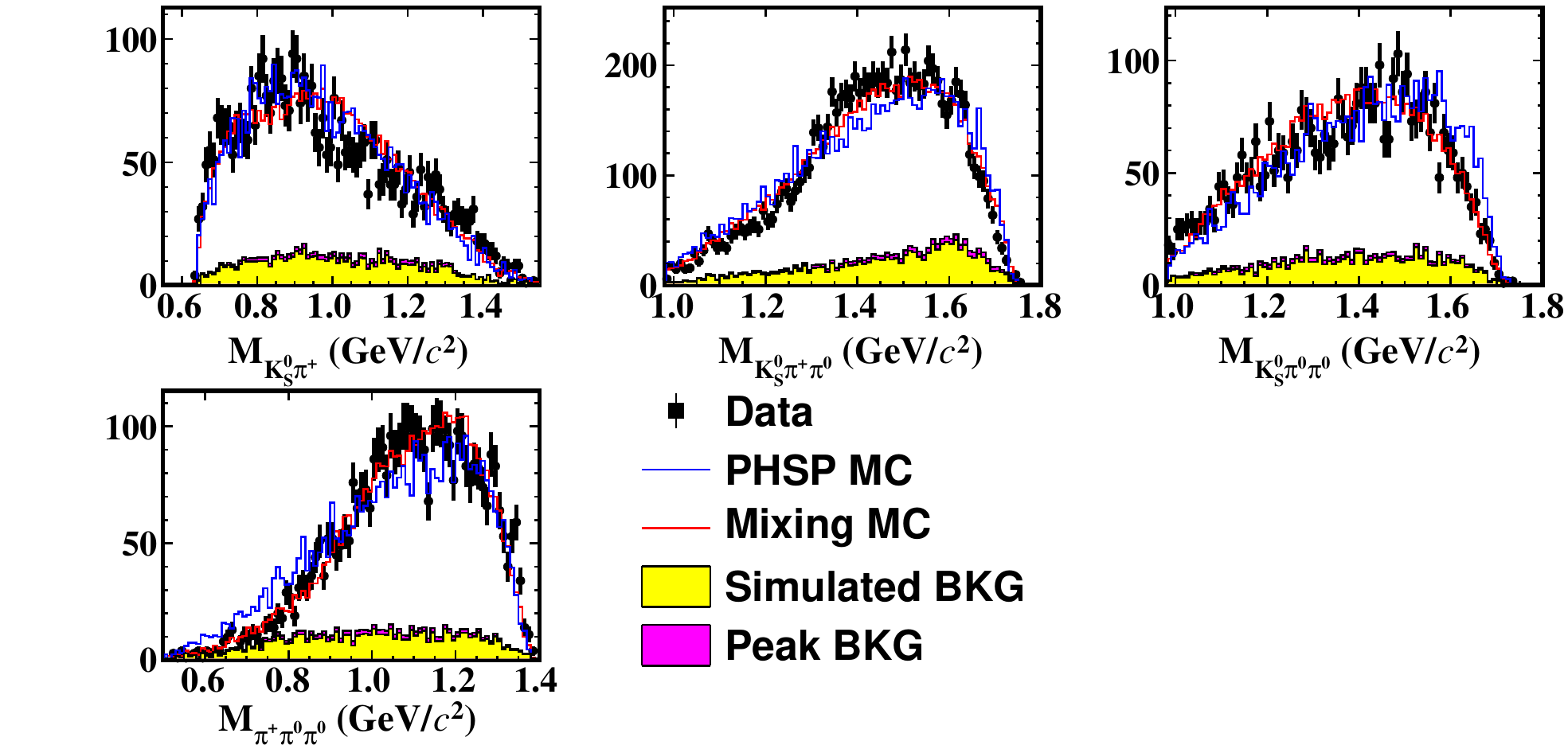}
  \caption{
  Comparisons of some typical distributions for the $D^+\to K^0_S\pi^+\pi^0\pi^0$ candidate events
  between data (dots with error bars) and the signal MC events (blue dotted/red dashed histograms)
  plus the MC-simulated backgrounds from the inclusive MC sample (yellow histograms).}
\label{fig:compare} 

\end{figure} 

In the fits, the background yields in the 2D $K^0_S$ sideband 1 and sideband 2 regions have been subtracted using Eq.~(\ref{eq:1}).
No significant signal of $D^0\to K^0_SK^0_S\pi^0$ is found.
The resulting upper limit on the branching fraction for $D^0\to K^0_S K^0_S\pi^0$ is $1.57\times 10^{-4}$ at a 90\% confidence level, using the Bayesian approach~\cite{UP_M} after incorporating the systematic uncertainty discussed in Sec.~\ref{sec:sys}.
The distribution of likelihood versus branching fraction is shown in Fig.~\ref{scan}.

Figure~\ref{fig:2Dfit} shows the $M^{\rm tag}_{\rm BC}$ and
$M^{\rm sig}_{\rm BC}$ projections of the 2D fits to data.
For the candidate events in the 2D $K^0_S$ sideband region, the 2D fits are performed similarly.
From these fits, we obtain the DT yields for the individual signal decays as shown in Table~\ref{tab:DT}.

 The double-tag efficiencies are determined from an MC simulation. To account for the effect of intermediate resonance structure on the efficiency, each of these decays is modeled by the corresponding mixed-signal MC samples, in which the dominant decay modes containing resonances of
$\eta$, $\omega$,
$K^*(892)$, $\rho(770)$, $f_0(980)$, $K_1(1270)$, and $K_1(1400)$ are mixed with the phase-space (PHSP) signal MC samples. The mixing ratios are determined by
examining the corresponding invariant mass and momentum spectra.
The momentum and the polar angle distributions of the daughter particles and the invariant masses of each two-, three- and four-body particle combinations
of the data agree with those of the MC simulations.
As an example, Fig.~\ref{fig:compare} shows
the momentum and the polar angle distributions of the daughter particles,
the invariant mass distributions of two- or three-body particle combinations of
the candidates for $D^+\to K^0_S\pi^+\pi^0\pi^0$ between data and MC simulations.

The measured values of $N^{\rm net}_{{\rm DT}}$,
$\epsilon^{}_{{\rm sig}}$, and the obtained branching fractions are summarized in Table~\ref{tab:DT}.
The signal efficiencies have been corrected by the data-MC differences in the selection efficiencies of $K^\pm$ and $\pi^\pm$ tracking and PID
procedures and $\pi^0$ reconstruction.
These efficiencies also include the branching fractions of the $K^0_S$ and $\pi^0$ decays.

\begin{table*}[htbp]
\centering
\caption{\small
Requirements of $\Delta E_{\rm sig}$,
the fitted and net yields of double-tag candidates ($N_{\rm DT}^{\rm fit}$ and $N^{\rm net}_{{\rm DT}}$),
background yield in the $K^0_S$ sideband ($N_{K^0_S,\rm sid}$), signal efficiencies ($\epsilon_{\rm sig}$),
and the obtained branching fractions (${\mathcal B}_{\rm sig}$) for various signal decays.
The first and second uncertainties for ${\mathcal B}_{\rm sig}$ are statistical and systematic, respectively,
while the uncertainties for $N^{\rm fit}_{\rm DT}$, $N_{K^0_S,\rm sid}$, $N^{\rm net}_{\rm DT}$ and $\epsilon_{\rm sig}$ are statistical only.
For $D^0\to K^0_SK^0_S\pi^0$, the $N^{\rm net}_{\rm DT}$ and ${\mathcal B}_{\rm sig}$ are set 90\% confidence level. The $N_{K^0_S,\rm sid}$ denotes the net $K^0_S$ background yield combined sideband 1 and sideband 2 regions.
}\label{tab:DT}
  \footnotesize
\begin{threeparttable}

\begin{tabular}{lccccccc}
  \hline\hline

  \multirow{2}{*}{Signal mode}&$\Delta E_{\rm sig}$ &\multirow{2}{*}{$N^{\rm fit}_{\rm DT}$ }&\multirow{2}{*}{$N_{K^0_S,\rm sid}$}  &\multirow{2}{*}{$N^{\rm net}_{\rm DT}$}  &$\epsilon_{\rm sig}$    &  ${\mathcal B}_{\rm sig}$\,  \\

  &(MeV) &&& &  \,(\%) &  ($10^{-3}$)  \\ \hline

$D^0\to K^0_S\pi^0\pi^0\pi^0$     &$(-73,34)$&\ \,$913\pm33$&$\ \,86\pm11$&$\ \,870\pm36$&$4.90\pm0.04$&$\ \,7.64\pm0.30\pm0.29$\\

$D^0\to K^-\pi^+\pi^0\pi^0\pi^0$  &$(-64,33)$&$1560\pm48$& --          &$1560\pm48$&$7.04\pm0.06$&$\ \,9.54\pm0.30\pm0.31$\\

$D^0\to K^0_S\pi^+\pi^-\pi^0\pi^0$&$(-50,30)$&$1253\pm40$&$134\pm14$&$1186\pm40$&$4.04\pm0.04$&$12.66\pm0.45\pm0.43$\\

$D^+\to K^0_S\pi^+\pi^0\pi^0$     &$(-63,34)$&$3513\pm66$&$226\pm19$&$3400\pm66$&$7.51\pm0.07$&$29.04\pm0.62\pm0.87$\\

$D^+\to K^0_S\pi^+\pi^+\pi^-\pi^0$&$(-45,30)$&$1097\pm37$&$107\pm14$&$1043\pm38$&$4.38\pm0.04$&$15.28\pm0.57\pm0.60$\\

$D^+\to K^0_S\pi^+\pi^0\pi^0\pi^0$&$(-43,25)$&$\ \,294\pm22$ &$ 19\pm 7$&$\ \,285\pm23$&$3.30\pm0.03$&$ \ \,5.54\pm0.44\pm0.32$\\

$D^+\to K^-\pi^+\pi^+\pi^0\pi^0$  &$(-54,31)$&$\ \,756\pm39$ & --          &$\ \,756\pm39$&$9.80\pm0.07$&$ \ \,4.95\pm0.26\pm0.19$\\
$D^0\to K^0_S K^0_S\pi^0$         &$(-45,28)$&$\ \,\ \,65\pm10$ &$118\pm13$&$6 \pm 13 (<24.6)$ &$ 7.06\pm0.11$&$<0.145$\\

  \hline\hline
\end{tabular}
\end{threeparttable}
\end{table*}
\section{Systematic uncertainties}
\label{sec:sys}

The systematic uncertainties are estimated relative to the measured branching fractions and are discussed below.
In the determination of the branching fractions using Eq.~(\ref{eq:br}), all uncertainties associated with the selection of tagged $\bar D$ are canceled.
The systematic uncertainties in the total yields of single-tag $\bar D$ mesons, which are mainly due to the fits to the $M_{\rm BC}$ distributions of the single-tag $\bar D$ candidates, were previously estimated to be
0.5\% for both neutral and charged $\bar D$~\cite{epjc76,cpc40,bes3-pimuv}.

The tracking and PID efficiencies for $K^\pm$ or $\pi^\pm$, $\epsilon_{K\,{\rm or}\,\pi}^{\rm tracking\,(PID)}[{\rm data}]$ and $\epsilon_{K\,{\rm or}\,\pi}^{\rm tracking\,(PID)}$  [{MC}],
are investigated using double-tag $D\bar D$ hadronic events.
The averaged ratios between data and MC efficiencies ($f_{K\,{\rm or}\,\pi}^{\rm tracking\,(PID)}=\epsilon_{K\,{\rm or}\,\pi}^{\rm tracking\,(PID)}[{\rm data}]/\epsilon_{K\,{\rm or}\,\pi}^{\rm tracking\,(PID)}[{\rm MC}]$) of tracking (PID) for $K^\pm$ or $\pi^\pm$ are weighted by the corresponding momentum spectra of signal MC events,
giving $f_K^{\rm tracking}$ ranging from $1.019-1.032$ and $f_\pi^{\rm tracking}$ close to unity for all seven signal modes. After correcting the MC efficiencies by $f_K^{\rm tracking}$,
the statistical uncertainties of $f_{K\,{\rm or}\,\pi}^{\rm tracking}$ are assigned as the systematic uncertainties of tracking efficiencies,
which are 0.2\% per $K^\pm$ and ($0.2-0.3$)\% per $\pi^\pm$. $f_K^{\rm PID}$ and $f_\pi^{\rm PID}$ are all close to unity and their individual uncertainties, ($0.2-0.3$)\%,
are taken as the associated systematic uncertainties per $K^\pm$ or $\pi^\pm$.

The systematic error related to the uncertainty in the $K_{S}^{0}$ reconstruction efficiency
is estimated from measurements of $J/\psi\to K^{*}(892)^{\mp}K^{\pm}$ and $J/\psi\to \phi K_S^{0}K^{\pm}\pi^{\mp}$ control samples~\cite{sysks}
and found to be 1.6\% per $K^0_S$.
The systematic uncertainty of $\pi^0$ reconstruction efficiency is assigned as ($0.7-0.8$)\% per $\pi^0$ from a study of double-tag $D\bar D$ hadronic decays of $\bar D^0\to K^+\pi^-\pi^0$ and $\bar D^0\to K^0_S\pi^0$ decays tagged by either $D^0\to K^-\pi^+$ or $D^0\to K^-\pi^+\pi^+\pi^-$~\cite{epjc76,cpc40}.
The systematic uncertainty in the 2D fit to the $M_{\rm BC}^{\rm tag}$ versus $M_{\rm BC}^{\rm sig}$ distribution is examined via the repeated measurements in which the signal shape ($\pm 1 \sigma$ in mean and width of smearing Gaussian)
and the endpoint of the ARGUS function ($\pm0.2$\,MeV/$c^2$) are varied.
Quadratically summing the changes of the branching fractions for these two sources yields the corresponding systematic uncertainties of ($0.8-4.9$)\%.

The systematic uncertainty due to the $\Delta E_{\rm sig}$ requirement is assigned to be (0.3-0.8)\% for various signal decays,
which corresponds to the largest efficiency difference with and without smearing the data-MC Gaussian resolution of $\Delta E_{\rm sig}$ for signal MC events.
Here, the smeared Gaussian parameters are obtained by using the samples of double-tag events $D^0\to K^0_S\pi^0$, $D^0\to K^-\pi^+\pi^0$, $D^0\to K^-\pi^+\pi^0\pi^0$, and $D^+\to K^-\pi^+\pi^+\pi^0$ versus the same $\bar D$ tags in our nominal analysis.
The systematic uncertainties due to $K^0_S$ sideband choice and $K^0_S$ rejection mass window
are cross checked by examining the changes of the branching fractions via varying nominal $K^0_S$ sideband and corresponding rejection window by $\pm5$~MeV/$c^2$. The shifts in the fitted results are negligible in the cross check and hence no further systematic uncertainty is considered.  
{\color{blue}{For the decays whose efficiencies are estimated with mixed signal MC events, the imperfect simulations of the momentum and $\cos\theta$ distributions of charged particles are considered as a source of systematic uncertainty listed as the MC modeling. To estimate this systematic uncertainty, we examine the change of the signal efficiency after removing the most significant mixed component except for the processes containing $\eta$ and $\omega$. In addition, for the decays involving $D\to \bar{K}\pi\eta$ and $D\to \bar{K}\pi\omega$, we vary the known branching fractions of $D\to \bar{K}\pi\eta$ and $D\to \bar{K}\pi\omega$ by $\pm 1\sigma$. For each signal decay, the quadratic sum of the efficiency changes is assigned as the corresponding systematic uncertainty. }}
The change of the re-weighted to nominal efficiencies, (0.3-2.9)\% for various signal decays, is assigned as the corresponding systematic uncertainty.

For $D^0\to K^0_S\pi^0\pi^0\pi^0$ and $D^0\to K^0_S\pi^+\pi^-\pi^0\pi^0$, after correcting the measured branching fractions by the QC factors, the residual uncertainties, 0.7\% and 0.6\%, are assigned as individual systematic uncertainties. The QC effect on $D^0\rightarrow K^-\pi^+\pi^0\pi^0\pi^0$ appearing through mixing and doubly-Cabibbo-suppressed decays is estimated by the method of Ref.~\cite{cleo-2Dfit}, which is controlled by the ratio of Cabibbo-suppressed and Cabibbo-favored rates combined with the strong phase difference between two amplitudes. The uncertainty is assigned to be 0.6\%.

The uncertainties due to the limited MC statistics for various signal decays, (0.6-0.9)\%, are taken into account as a systematic uncertainty. The uncertainties of the quoted branching fractions of the $K^0_S\to \pi^+\pi^-$ and
$\pi^0\to \gamma\gamma$ decays are 0.07\% and 0.03\%, respectively~\cite{pdg2020}.

Table~\ref{tab:relsysuncertainties1} summarizes the systematic uncertainties in the branching fraction measurements. For each signal channel, the total systematic uncertainty is obtained by adding the above sources quadratically. The obtaned total systematic systematic uncertainties are in the range of (3.0-5.8)\% for various signal modes.

\begin{table*}[htbp]
\centering
\caption{
Relative systematic uncertainties (\%) in the branching fraction measurements of the signal decays
(1) $D^0\to K^0_S\pi^0\pi^0\pi^0$,
(2) $D^0\to K^-\pi^+\pi^0\pi^0\pi^0$,
(3) $D^0\to K^0_S\pi^+\pi^-\pi^0\pi^0$,
(4) $D^+\to K^0_S\pi^+\pi^0\pi^0$,
(5) $D^+\to K^0_S\pi^+\pi^+\pi^-\pi^0$,
(6) $D^+\to K^0_S\pi^+\pi^0\pi^0\pi^0$,
(7) $D^+\to K^-\pi^+\pi^+\pi^0\pi^0$, and
(8) $D^0\to K^0_S K^0_S\pi^0$. Uncertainties which are not applicable are denoted by ``--''. }
\label{tab:relsysuncertainties1}
\centering
\begin{tabular}{clllllllll}
\hline\hline

Source&\ 1&\ 2&\ 3&\ 4&\ 5&\ 6&\ 7&\ 8\\ \hline
$N^{\rm tot}_{\rm ST}$           &0.5 &0.5 &0.5 &0.5 &0.5 &0.5 &0.5&0.5 \\
$(K/\pi)^\pm$ tracking           &\,\,-- &0.4 &0.4 &0.2 &0.6 &0.2 &0.6&\,\,-- \\
$(K/\pi)^\pm$ PID                &\,\,-- &0.4 &0.4 &0.2 &0.6 &0.2 &0.6&\,\,-- \\
$K^0_S$ reconstruction           &1.6 &1.6 &\,\,-- &1.6 &1.6 &1.6 &\,\,--&3.2 \\
$\pi^0$ reconstruction           &2.1 &2.1 &1.4 &1.4 &0.7 &2.1 &1.4&0.7 \\
2D fit                           &1.8  &2.0  &1.9  &0.8  &1.6 &4.9 &3.5&3.9 \\
$\Delta E_{\rm sig}$ requirement &0.3 &0.6 &0.5 &0.7 &0.4 &0.8 &0.6&0.6 \\
Quoted $\mathcal B$              &0.12 &0.10 &0.10 &0.10 &0.08 &0.12 &0.07&0.14 \\
MC modeling                      &1.8  &0.6  &1.2  &1.6  &2.9 &0.8  &0.3&\,\,-- \\
MC statistics                    &0.7  &0.8  &0.9  &0.6  &0.8 &0.9  &0.6&0.5 \\
QC effect                        &0.7  &0.6  &0.6  &\,\,-- &\,\,-- &\,\,-- &\,\,--&0.7 \\
\hline
Total                            &3.8  &3.3  &3.4  &3.0  &4.0 &5.8 &4.0&5.2 \\
\hline\hline
\end{tabular}
\end{table*}

\section{Summary}

In summary, we present the first measurements of the branching fractions of
the hadronic decays of
$D^0\to K^0_S\pi^0\pi^0\pi^0$,
$D^0\to K^-\pi^+\pi^0\pi^0\pi^0$,
$D^0\to K^0_S\pi^+\pi^-\pi^0\pi^0$,
$D^+\to K^0_S\pi^+\pi^0\pi^0$,
$D^+\to K^0_S\pi^+\pi^0\pi^0\pi^0$,
$D^+\to K^-\pi^+\pi^+\pi^0\pi^0$, and
$D^+\to K^0_S\pi^+\pi^+\pi^-\pi^0$.
After subtracting the known branching fractions  $\mathcal B^{\rm prd}_{D^{0(+)}\rightarrow \bar{K}\pi\eta}$ and $\mathcal B^{\rm prd}_{D^{0(+)}\rightarrow \bar{K}\pi\omega}$ for $D\to  \bar{K}\pi\eta$ and $D\to  \bar{K}\pi\omega$ From these decays,
the residual branching fractions are summarized in the last columns of Table \ref{tab:final_result}. Here, $\mathcal B^{\rm prd}_{D^{0(+)}\rightarrow  \bar{K}\pi\eta}  =\mathcal B_{D^{0(+)}\rightarrow  \bar{K}\pi\eta}  \times \mathcal B_{\eta\to 3\pi}$,
$\mathcal B^{\rm prd}_{D^{0(+)}\rightarrow  \bar{K}\pi\omega}=\mathcal B_{D^{0(+)}\rightarrow  \bar{K}\pi\omega}\times \mathcal B_{\omega\to 3\pi}$, and $\bar{K}$ denotes $K^0_S$ when a $K^0_S$ meson is involved in the decay.
Except for the decay $D^{+}\to K^{0}_{S}\pi^{+}\pi^{0}\pi^{0}\pi^{0}$,
significant non-($\eta,\omega$) contributions have been found.
In the near future, further amplitude analyses of these decays with larger data samples at BESIII~\cite{bes3-white-paper} and Belle~II~\cite{belle2-white-paper} will provide rich information about the multi-body hadronic $D$ decays to scalar, vector, axial and tensor mesons, which will benefit further understanding of quark SU(3)-flavor symmetry.

\begin{table*}[htbp]
\centering
\caption{
Obtained BFs ($\mathcal B_{\rm sig}$, $\mathcal B^{\rm prd}_{D^{0(+)}\rightarrow  \bar{K}\pi\eta}$, $\mathcal B^{\rm prd}_{D^{0(+)}\rightarrow  \bar{K}\pi\omega}$ and $\mathcal B_{\rm non{\text -}\eta,\omega}=\mathcal B_{\rm sig}-\mathcal B^{\rm prd}_{D^{0(+)}\rightarrow  \bar{K}\pi\eta}-\mathcal B^{\rm prd}_{D^{0(+)}\rightarrow  \bar{K}\pi\eta}$) for various signal decays (in units of $ 10^{-3}$).
}
\label{tab:final_result}
\centering

\begin{tabular}{lcccc}
  \hline\hline
Decay mode & $\mathcal B_{\rm sig}$ & $\mathcal B^{\rm prd}_{D^{0(+)}\rightarrow  \bar{K}\pi\eta}$ & $\mathcal B^{\rm prd}_{D^{0(+)}\rightarrow  \bar{K}\pi\omega}$ & $\mathcal B_{\rm non{\text -}\eta,\omega}$\\  \hline
$D^0\to K^{0}_{S}\pi^{0}\pi^{0}\pi^{0}$             &  $ \ \,7.64\pm  0.30\pm  0.29$  &  $ 1.66\pm  0.04$  & --           &$\ \,  5.98\pm  0.30\pm  0.29$ \\
$D^0\to K^{-}\pi^{+}\pi^{0}\pi^{0}\pi^{0}$          &  $ \ \, 9.54\pm  0.30\pm  0.31$  &  $ 6.06\pm  0.13$  & --          &$\ \,   3.48\pm  0.30\pm  0.34$\\
$D^0\to K^{0}_{S}\pi^{+}\pi^{-}\pi^{0}\pi^{0}$      &  $ 12.66\pm  0.45\pm  0.43$  &  $ 2.31\pm  0.11$  &  $ 7.14\pm  0.47$&$\ \, 3.21\pm  0.45\pm  0.65$ \\
$D^{+}\to K^{0}_{S}\pi^{+}\pi^{0}\pi^{0}$           &  $ 29.04\pm  0.62\pm  0.87$  &  -- &   --        &  $ 29.04\pm 0.62\pm  0.87$ \\
$D^{+}\to K^{0}_{S}\pi^{+}\pi^{+}\pi^{-}\pi^{0}$    &  $ 15.28\pm  0.57\pm  0.60$  &  $ 3.00\pm  0.11$  &  $ 6.29\pm  0.44$    &$\ \,  5.99\pm  0.57\pm  0.75$\\
$D^{+}\to K^{0}_{S}\pi^{+}\pi^{0}\pi^{0}\pi^{0}$    &  $ \ \, 5.54\pm  0.44\pm  0.32$  &  $ 4.28\pm  0.16$  & --      &$\ \,  1.26\pm 0.44\pm  0.36$ \\
$D^{+}\to K^{-}\pi^{+}\pi^{+}\pi^{0}\pi^{0}$        &  $ \ \, 4.95\pm  0.26\pm  0.19$  &  -- &   --       &$\ \,  4.95\pm  0.26\pm  0.19$ \\  \hline\hline
\end{tabular}
\end{table*}

\section{Acknowledgement}
The BESIII collaboration thanks the staff of BEPCII and the IHEP computing center for their strong support. This work is supported in part by National Key R\&D Program of China under Grants No. 2020YFA0406400 and No. 2020YFA0406300; National Natural Science Foundation of China (NSFC) under Grants No. 11875170, No. 12035009, No. 11625523, No. 11635010, No. 11735014, No. 11822506, No. 11835012, No. 11935015, No. 11935016, No. 11935018, No. 11961141012, No. 12022510, No. 12025502, No. 12035013, and No. 12061131003, No. 12192260, No. 12192261, No. 12192262, No. 12192263, No. 12192264, No. 12192265; the Chinese Academy of Sciences (CAS) Large-Scale Scientific Facility Program; Joint Large-Scale Scientific Facility Funds of the NSFC and CAS under Grants No. U1832207 and No. U1732263; CAS Key Research Program of Frontier Sciences under Grant No. QYZDJ-SSW-SLH040; 100 Talents Program of CAS; The Institute of Nuclear and Particle Physics (INPAC) and Shanghai Key Laboratory for Particle Physics and Cosmology; ERC under Grant No. 758462; European Union's Horizon 2020 research and innovation programme under Marie Sklodowska-Curie grant agreement under Grant No. 894790; German Research Foundation DFG under Grant No. 443159800, Collaborative Research Center CRC 1044, FOR 2359, FOR 2359, GRK 2149; Istituto Nazionale di Fisica Nucleare, Italy; Ministry of Development of Turkey under Grant No. DPT2006K-120470; National Science and Technology fund; Olle Engkvist Foundation under Grant No. 200-0605; STFC (United Kingdom); The Knut and Alice Wallenberg Foundation (Sweden) under Grant No. 2016.0157; The Royal Society, UK under Grants No. DH140054 and No. DH160214; The Swedish Research Council; U. S. Department of Energy under Grants No. DE-FG02-05ER41374 and No. DE-SC-0012069.

\end{document}